\documentclass[12pt,twocolumn]{article}
\usepackage{tabularx}
\usepackage[english]{babel}
\usepackage[utf8x]{inputenc}
\usepackage[T1]{fontenc}

\usepackage{setspace}

\usepackage{enumitem}

\usepackage[twocolumn,top=1cm,bottom=1cm,left=1cm,right=1cm,marginparwidth=1cm]{geometry}
\usepackage{geometry}

\usepackage{amsmath}
\usepackage{graphicx}
\usepackage[colorinlistoftodos]{todonotes}
\usepackage[colorlinks=true, allcolors=blue]{hyperref}
\usepackage{authblk}

\providecommand{\keywords}[1]
{
  \immediate	
  \textbf{\textit{Keywords-}} #1
}

\title{Detecting Phishing Sites - An Overview}

\date{}

\makeatletter
\newcommand\level[1]{%
  \ifcase#1\relax\expandafter\chapter\or
    \expandafter\section\or
    \expandafter\subsection\or
    \expandafter\subsubsection\else
    \def\next{\@level{#1}}\expandafter\next
  \fi}
\newcommand{\@level}[1]{%
  \@startsection{level#1}
    {#1}
    {\z@}%
    {-3.25ex\@plus -1ex \@minus -.2ex}%
    {1.5ex \@plus .2ex}%
    {\normalfont\normalsize\bfseries}}

\newcounter{level4}[subsubsection]
\@namedef{thelevel4}{\thesubsubsection.\arabic{level4}}
\@namedef{level4mark}#1{}
\count@=4
\loop\ifnum\count@<100
  \begingroup\edef\x{\endgroup
    \noexpand\newcounter{level\number\numexpr\count@+1\relax}[level\number\count@]
    \noexpand\@namedef{thelevel\number\numexpr\count@+1\relax}{%
      \noexpand\@nameuse{thelevel\number\count@}.\noexpand\arabic{level\number\numexpr\count@+1\relax}}
    \noexpand\@namedef{level\number\numexpr\count@+1\relax mark}####1{}}
  \x
  \advance\count@\@ne
\repeat
\makeatother
\author{P.Kalaharsha$^{a,b}$ , B. M. Mehtre$^{a}$  \\
        \small $^{a}$Center of excellence in cyber security, Institute for Development and Research in Banking Technology (IDRBT), Hyderabad, India \\
        \small $^{b}$School of Computer Science and Information Sciences
(SCIS), University of Hyderabad, Hyderabad, India\\  } 

\begin{document}

\maketitle{}

\begin{abstract}
Phishing is one of the most severe cyber-attacks where researchers are interested to find a solution. In phishing, attackers lure end-users and steal their personal information. To minimize the damage caused by phishing must be detected as early as possible. There are various phishing attacks like spear phishing, whaling, vishing, smishing, pharming and so on. There are various phishing detection techniques based on white-list, black-list, content-based, URL-based, visual-similarity and machine-learning. In this paper, we discuss various kinds of phishing attacks, attack vectors and detection techniques for detecting the phishing sites. Performance comparison of 18 different models along with nine different sources of datasets are given. Challenges in phishing detection techniques are also given.
\end{abstract}


\section{Introduction}
In recent days cyber-attacks are increasing at an unprecedented rate. Phishing is one among those cyber-attacks [55]. In phishing, attackers lure the end-users by making them click the hyper-links which make them lose their personally identifiable information, banking and credit card details, and passwords. In this attack the attackers disguise themselves as trusted entities such as service providers, employees of the organization or technical-support team from the organization so that end-users never doubt them. It is mainly done through emails asking to update the system, or saying that account has been suspended, or asking to claim the prize and so on [59]. The main goal of phishing is to make end-users share their sensitive information. Now-a-days information regarding anything is available online and that information is stored in websites. Websites help the end-users by providing them information about their respective products,services or helping the end-users if they face any problem by chat-bots, message forums and so on. Websites also store the personal information of the end-users. As websites help the end-users in gaining information they can be used as bait for trapping the end-users to obtain confidential information from them. Websites can be forged easily with the help of many online tools. The forged websites look exactly similar to the legitimate websites and end-users will not get any doubt while browsing these sites. The illegitimate websites must be detected as early as possible to make sure that there is no loss of information.
\subsection{Types of Phishing Attacks}
There are different types of Phishing attacks. The main goal of these attacks are to obtain sensitive information from the end-users.
\begin{figure}[htpb!] 
    \centering 
\includegraphics[scale=0.5]{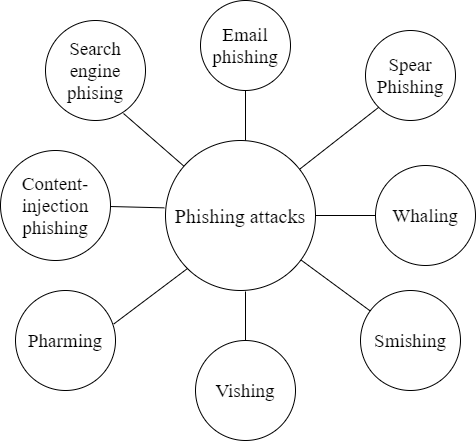} 
    \caption{Types Of Phishing Attacks} 
    \label{fig:label1} 
\end{figure}
Figure 1 shows various types of phishing attacks.
\subsubsection{ Email Phishing}
In this type of phishing an attacker sends an email regarding any problem, update or any sensitive matter that must be changed immediately once the user clicks the email and all the input the details entered by the end-users will be redirected to the attacker [56].
\subsubsection{Spear Phishing}
In this attackers aims for specific individuals or enterprises, as opposed to random application users. It’s a more in-depth version of phishing that requires special knowledge about an organization, including its power structure. In this attack emails are sent to specific persons unlike phishing [56].
\subsubsection{Whaling}
It is known as whaling phishing or a whaling phishing attack. It is a form of spear phishing where, in this phishing attackers target high-profile employees, such as the CEO or CFO, in order to steal sensitive information from a company. As these people hold higher positions within the company, they will have complete access to sensitive data. It will be easy to obtain more information [57].
\subsubsection{Smishing}
It is also known as SMS phishing. It is a type of social engineering attack carried out in order to steal user data including personal information, financial information, and credentials. Smishing also aims at laundering money from victims. In Smishing, scammers send phishing messages via an SMS text that includes a malicious link. The phishing messages trick recipients into clicking the malicious link, which redirects them to a phishing page where personal information is harvested [56][58].\newline
Example : Lucky Draw campaign - In this attackers send SMS to end-users asking them to claim the account that they have won through lucky draw. Attackers ask them to click the link and provide their information so that amount will be transferred to their accounts.
\subsubsection{Vishing}
It is also known as Voice Phishing. It is a type of phone fraud that uses voice messages to obtain personal information or money from victims. Vishing uses automated voice recordings to lure victims. In Vishing, an automated voice call stating that the recipients’ bank account has been compromised is sent. The voice message then asks the recipient to call a specified toll-free number. Once users call to that toll-free number, the user’s bank account number and other personal details are harvested via the phone keypad [56][58].
\subsubsection{Pharming}
Pharming is sometimes known as “phishing without a lure”. When a user attempts to navigate to a site, their computer can determine the IP address by either consulting a local file of defined mappings—a hosts file—or by consulting a DNS server on the Internet. Pharming is usually conducted either by changing the hosts file on a victim’s computer (hosts file pharming) or by exploiting a vulnerability in DNS server software (DNS poisoning) [60].
\subsubsection{Content-injection Phishing}
In this, the content of the legitimate website is replaced with some random content  with different input fields similar to legitimate site so that end –users trust easily and give their data easily [60].
\subsubsection{Search Engine Phishing}
It occurs when phishers create websites with attractive sounding offers and have them indexed legitimately with search engines. Users find these sites in the normal course of searching for products or services and are fooled into giving up their information [60].
\subsection{Phishing techniques}
There are different techniques used by attackers to execute different types of phishing attacks. By using these techniques the attackers can bypass the security and are able to obtain confidential information from the end-users [61].\newline
1. Link Manipulation \newline
2. Website Forgery \newline
3. Pop ups
\subsubsection{Link Manipulation}
Link manipulation [61] is a widely used technique for phishing scams. It is done by directing a user through fraud to click a link to a fake website. Hackers are now using manipulative ways to get the users to click such as:\newline
              1.Use of sub-domains\newline
              2.Hidden URLs\newline
              3.Misspelled URLs\newline
              4.IDN(internationalized domain name) Homograph attacks
\subsubsection{Website Forgery}
Website forgery [61] is another phishing technique that works by making a malicious website impersonate an authentic one, so as to make the visitors give up their sensitive information like account details, passwords, credit card numbers, etc. Web forgery is mainly carried out in two ways: \newline
1. Cross-site scripting \newline
2. Website spoofing.
\subsubsection{Pop-up}
Pop-up messages are one of the easiest techniques to conduct successful phishing scams. They allow hackers to steal login details by sending users pop-up messages and eventually leading them to forged websites through these pop-ups. A variant of phishing attacks, also known as “in-session phishing,” works by displaying a pop-up window during an online banking session and appears to be a message from the bank [61].
\subsection{Phishing Websites}
Phishing websites are replicas of legitimate websites. For phishing websites,
entire website is not created, only the home page or page where user can give inputs is created so that any information entered is sent to the attacker. There are many ways to create a phishing website such as downloading source code of any particular website, clone the website or use any specific tools. In kali-linux there is a tool called SET (social-engineering toolkit) which is mainly used for cloning of web-pages [70]. The difference between the
phishing websites and legitimate websites can be seen in their URLs, content of the websites, logos seen on the websites, hyperlinks, hosting domains, domain age, source-code, SSL certificates [43] etc. Now-a-days as URLs being too large,they are being shortened. These are done by URL shorteners [50]. These URLs look different from normal URLs, they don't contain any domain, sub-domain,Top-level domain(TLD), they just contain protocol and path. These URLs just redirects to the original URL. In this method as there is no information regarding about website, attackers make use of this method and send this type of URLs to the end-users. End-users doesn't pay much attention to URL, they just click the URL and enter the credentials, in that way they lose their confidential information. These websites must be identified as early as possible to prevent loss of data.\newline\newline
The rest of the paper is organized into different sections: Section 2 lists various website detection methods over time, Section 3 describes about different sources of datasets, comparison between different models along with their accuracy and different challenges to be worked on , Section 4 presents challenges and discussions, and Section 5 presents the conclusion followed by references.

\section{ Phishing Website Detection Techniques}
\begin{figure*}[htpb!] 
    \centering 
\includegraphics[width=18cm, height=6cm]{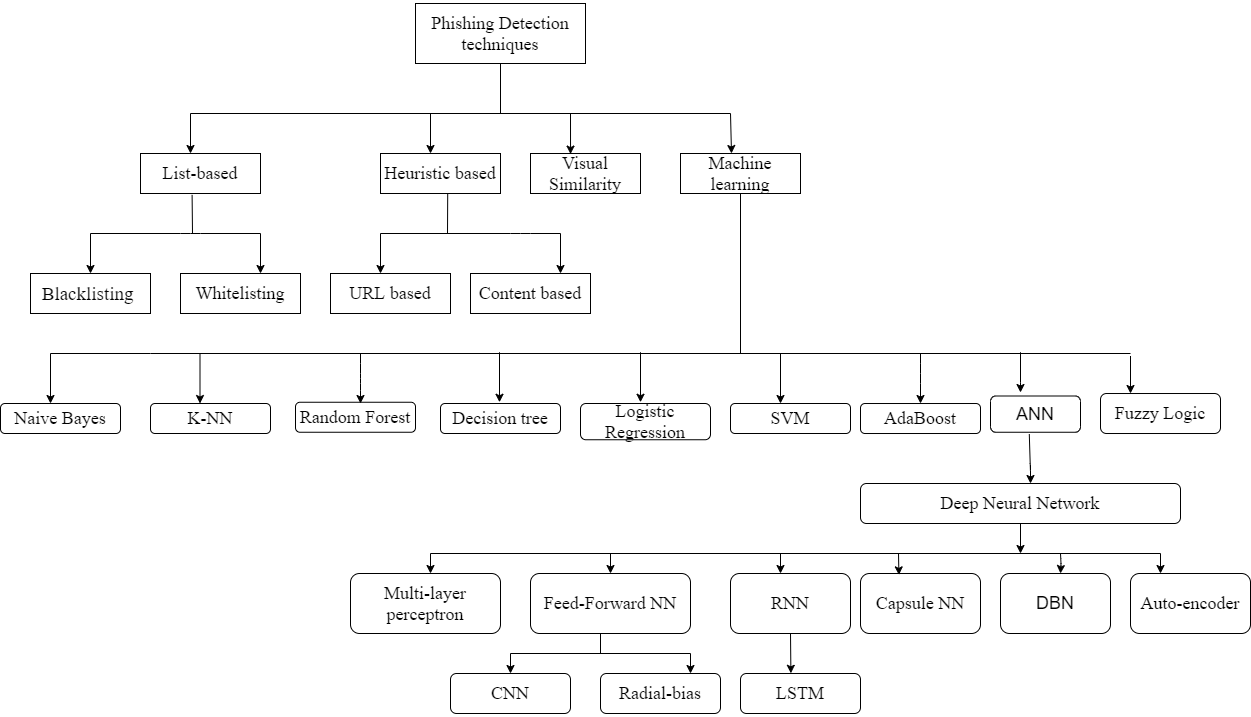} 
    \caption{Phishing Website-Detection Techniques} 
    \label{fig:label2} 
\end{figure*}
There are different types of detection techniques for detecting the phishing sites. Figure 2 shows various types of phishing detection techniques.
\subsection{List based}
In this method there are two types:\newline
1)Blacklisting \newline
2)White-listing.\newline
These are also known as traditional approaches or database-oriented approaches. Their response time and detection accuracy is very high [36][42].
\subsubsection{Black-Listing}
In this technique, the URL’s which are considered as phished sites are stored in database so when a new URL is entered it compares with the URLs in the database and if it matches it is blocked by the browser and is stored in the database for future purpose. The limitation of this technique is that detection of zero-hour phishing attacks is not possible [42].
\subsubsection{White-Listing}
In this technique the legitimate URLs are stored in the database and are used for checking new URLs. In this technique when a new URL is entered, first they check for that URL in the database, if there is no record of that URL then the entire information of that URL is checked such as domain names, age, SSL certificates, hyper-links connected to the website and then they are stored in the database. The drawback in case of white-listing technique is, are the websites registered as legitimate are really legitimate or they present themselves as legitimate. The limitation of listing techniques is they require large amount of space [42].
\subsection{Heuristic Oriented Detection}
It is extension of listing technique. In this technique, features of the websites are extracted such as URL’s, content and they are used for comparison among different sites. If they match then those new websites are considered as phishing sites. These are better than listing techniques and their results gives more accuracy but their response time is low [5][42][43][12]. A different approach for detection of zero-hour phishing attack is discussed in [37].
\subsubsection{URL-based}
For increasing the speed of the detection, URL based detection mechanisms are more popular. If URL-based features combined with machine learning gives better accuracy-rate [7][14][35].
\subsubsection{Content-based}
In this, the content of the websites are compared with the legitimate websites to determine whether the sites are legitimate or not [12]. But there are many websites which doesn't have much content then this detection strategy fails . Now-a-days website contents are replaced by images [36].
\subsection{Visual Similarity}
In this technique screenshots of web pages are taken and are stored in the databases. Then if there are look alike websites then the screenshots of both the web-pages are compared to detect if it is phishing website or not. The limitations of this technique are it consumes a longer execution time leading to be unrealistic. It requires large storage space for storing the screenshots of websites. When multiple websites with same URL appear the first one to appear is considered as legitimate. But there can be any chance that the first to appear can be a phishing site [27][36][42][54].
\subsection{Machine learning}
Based on different datasets obtained from the different features of the websites, machine learning trains model from those datasets and tests them with different machine learning classifiers such as Random forest classifier, Support vector machine, Decision Tree, Naive Bayes, Logistic regression and so on [32]. These classifiers help in predicting the websites even before they are created, thus machine learning solves the problem of zero-hour phishing attacks. The classifier’s accuracy varies depending on the size of the data-set and type of features used [2][26][45][48]. Frameworks are also developed to detect phishing attacks [51][53]. 
\subsubsection{Naive-Bayes}
Naive Bayes classifier is a generative probabilistic model in machine learning and is based on the Bayes theorem. It is mostly used in classification areas, such as text classification, spam detection, because of its simplicity. Its
features are independent among each other. Detection of websites using lexical analysis of URL in which Naive Bayes achieved the highest accuracy [6][16][52].
\subsubsection{Random Forest}
Random Forest is an ensemble classifier used for classification and regression. It constructs decision trees based on randomly selected sets in training samples and then aggregate decisions from these trees by averaging or majority voting. It improves accuracy and also reduces over-fitting [4][9][16].
\subsubsection{Decision Tree}
Decision tree classifiers are one of the most popular classifiers used in classification and regression. It divides the training data-set until it reaches to a leaf node, which is a label in classification. Decision tree classifier uses the entire training data-set while constructing a tree unlike
Random Forest [9][16][52].
\subsubsection{Support Vector Machine}
Support vector machine can solve linear or non-linear problems. In linear problems, it simply finds a hyper-plane in N-dimensional feature space [52]. Different versions of support vector machine can also be used for detection purposes [5][16][53]. 
\subsubsection{K-Nearest Neighbour}
KNN is a non-parametric algorithm used in both classification and regression. Its classification works on unknown data closest to k in the training feature space. Closest points are selected using distance functions such
as Hamming, Euclidian and Minkowski. KNN works slow if the data size is large [16].
\subsubsection{Logistic Regresssion}
Logistic regression is a discriminative probabilistic model mainly used in which the output is binary. Logistic regression performs better than Naive Bayes model when training size is close to infinity [16][52].
\subsubsection{AdaBoost}
AdaBoost (Adaptive Boosting) works as a conjunction algorithm because it is used to classify by training different weak learning algorithms to form a strong one i.e. to improve performance. The output of weak classifiers are combined by setting correct weights for final decision. Since AdaBoost is sensitive to outliers and focuses on hard-to-classify samples, it is less resistant to overfitting [21][46][52].
\subsubsection{Artifical Neural Networks}
Artificial neural networks(ANN) consist of information processing elements known to mimic neurons of the brain.ANN is classified into two types depending on the number of layers.they are shallow and deep neural networks. If the number of layers is two i.e, only input and output layer then it is shallow neural networks. If the neural networks consists of three layers in which there is at least one hidden layer then it is deep neural network [11][16][49][69].
\level{4}{Deep Neural Network}
Deep neural network represents the type of machine learning when the system uses many layers of nodes to derive high-level functions from input information. It means transforming the data into a more creative and abstract component [25][77]. To detect the problem of over-fitting, a new method has been proposed namely, OFS-NN, an effective phishing websites detection model based
on the optimal feature selection method and neural network. The proposed algorithm is able to alleviate the over-fitting problem of the underlying neural
network to a large extent [40].
\level{5}{Feed-Forward Neural Network}
A feed-forward neural network is an artificial neural network wherein connections between the nodes do not form a cycles. In this network, information travels from the input nodes, through the hidden nodes (if any) and to the output nodes in just one direction, forward. The network does not have any cycles or loops [13][71].
\level{6}{Convolutional Neural Networks}
Convolutional Neural Networks(CNN) belongs to the family of Artificial Neural Networks. A CNN is a deep learning technique that works well for identifying simple patterns in the data which will then be used to form more complex patterns in subsequent layers. It gives high accuracy when compared to other machine learning classifiers [8][10]. PhishingNet, a deep learning-based approach for timely detection of phishing Uniform Resource Locators (URLs) which uses CNN as well as RNN to extract the features [3][17][19][24][27][33][41][52].
\level{6}{Radial basis Function}
Radial Basis Function(RBF) is a type of feed forward neural network that has been used broadly for classification and regression problems because of its simplest architecture and effective results for numerical data [20].
\level{5}{Multi-layer perceptron}
Multilayer Perceptron (MLP) is a successful model in the field of deep learning.
It is a class of feed-forward supervised learning techniques. It has multiple layers of perceptron with a non-linear activation function rather than a single-layer perceptron; that's why it is called multiple layer perceptron. It uses a back propagation algorithm for supervised learning [13][22].
\level{5}{Recurrent Neural Networks}
A class of artificial neural networks where connections between nodes form a directed graph along a temporal sequence is recurrent neural network (RNN). This allows it to exhibit temporal dynamic behavior. Derived from feedforward neural networks, RNNs can use their internal state (memory) to process variable length sequences of inputs [15][18][41][72].
\level{6}{Long term Short memory}
The artificial recurrent neural network (RNN) architecture[1] used in the field of deep learning is long short-term memory (LSTM). LSTM has feedback links, unlike normal feedforward neural networks. It can process not only single data points (such as images), but also entire data sequences, such as speech or video [15][18][33][52][74].
\level{5}{Capsule Neural Network}
The machine learning system that is a type of artificial neural network (ANN) that can be used to better model hierarchical relationships is Capsule Neural Network (CapsNet). Add structures to a convolution neural network (CNN) called "capsules" and to reuse output from some of those capsules to form more stable representations for higher capsules is their idea. The output is a vector consisting of the likelihood of an observation and a pose for that observation[31][73]
\level{5}{Deep Belief Network}
Deep Belief Network(DBN) is a generative graphical model composed of multiple hidden layers with connection between layers and, there is no connection between units within each layer. DBN involves two steps that is, training in unsupervised way and training in supervised way. In unsupervised way, DBN learns to probabilistically reconstruct to input, this layer is called feature detectors on input. In supervised way, DBN works as classifier and does classification. DBN extracts the deep hierarchical representation (knowledge) and learns from this knowledge to make best model [29].
\level{5}{Auto-encoder}
Auto-Encoder(AE) is an unsupervised deep learning method. The basic framework of AE comprises an input layer, an output layer, and a hidden layer. Therein, the input layer and the output layer have the same structure, and when the input is
equal to the output, the hidden layer represents potential structure and characteristics of the input. The aim of AE is to transform inputs into outputs with the least possible amount of deviation [30].
\subsubsection{Fuzzy logic}Machine learning detects the sites through feature extraction but what types of features are considered, does the features considered are enough to detect the phishing sites and are those features present in all the sites. These can be known by using Fuzzy rough set, which is used as a tool to select the most effective features for detecting phishing websites by using most commonly used features such as domain based, address-bar, abnormal-based and HTML/JavaScript based [23][38][44][47].
\section{ Datasets \& Performance comparison}
In this section performance comparison of different models along with different datasets are discussed.
Datasets play an important role in predicting illegitimate websites. In techniques like Machine learning, deep learning and neural networks: first data will be collected from the different sources,then the collected data will be used for training the model and then the model will be used for testing to check whether it can predict the websites correctly or not. Some of the common sources of datasets are shown below in Table 1.
\begin{table}[!htpb]
\small
    \caption{Commondatasets}
    \begin{tabular}{ | m{0.07\linewidth} | m{0.3\linewidth} | m{0.2\linewidth}| m{0.3\linewidth} |}
\hline
S.No & Name of the datasets & Size of data-set & Description\\
\hline
1 & Alexa database[62]  & 1 million URLs  & Top 1 million sites\\
\hline
2 & Common-Crawl [63] & 940 million URLs& 2.8 billion webpages\\
\hline
3& Phish-tank  [64] & 68,40,198 URL's  & updated-daily \\
\hline
4&Open-Fish  [65] & 4,253 URLs & updated-daily\\
\hline
5 & UCI Machine learning repository [66] & 11,055 URL feature values & collected from 2456 different websites\\
\hline
6 & Majestic  [67] & 1 million URLs  & Top 1 million website in the world \\
\hline
7 & Ebbu2017  [68] & 73,575 URLs & both legitimate \& phishing URLs\\
\hline
8 & Kaggle [75] &11,000 URL'S & contains different datasets \\
\hline
9 & 5000 Best Websites[76] & 5000 URLs & Information about website is provided \\
\hline
\end{tabular}
\end{table}
Alexa and Common crawl contains names of the legitimate sites which are likely to be used for phishing [62][63]. Phish-tank,Open-Fish are the sites where end-users report the suspicious URL's to know whether they are phishing sites or not [64][65]. In UCI-Machine learning repository the data-set is collected from different sources and stored. This data-set is mainly used for research purposes [66]. In Majestic,the data-set contains domains with the referring subnets [67]. Ebbu2017 dataset is used in [39] to train the model. The data-set contains 73,575 URLs, out of which 36,400 are legitimate and 37,175 are phishing. Kaggle is an online repository which contains different datasets collected from different sources [75]. These datasets are helpful in training the models.\newline
Performance results from different papers are collected. These results differ based on the datasets used, features extracted from the websites, algorithms used and the different classifiers used for comparison.These all factors play a vital role in determining the accuracy. The results are shown in the Table 2.
\begin{table*}[!htpb]
\small
      \caption{Performance comparison of different techniques}
\begin{tabular}{ | m{0.04\linewidth} | m{0.4\linewidth}| m{0.4\linewidth} | m{0.1\linewidth} |} 
\hline
S.No & Detection technique &Source of the datasets & Accuracy \\ 
\hline
1 & Naive-Bayes[6] &Phish-tank,OpenFish,Majestic  &  97.18\\ 
\hline
2 & CNN [3] & Common-Crawl,Phish-tank& 96\\
\hline
3 & Random-forest [9] & Kaggle & 97 \\
\hline
4 & CNN [10] &UCI Machine Learning Repository&  97.3\\ 
\hline
5 & CNN+RNN[41] &Phish-tank,Open Phish,Alexa & 97.9\\ 
\hline
6 & DeepReinforcement Learning[39] & Ebbu2017Phishing Dataset  & 90.1 \\
\hline
7 & Character-level CNN[8]& Alexa,Openphish, spamhaus.org, techhelplist.com, isc.sans.edu\&phishtank & 95.02\\
\hline
8 &OFS-NN[40] & UCI Machine Learning Repository, Phishtank,Alexa &  99.3\\
\hline
9 & XCS [45] & Alexa,Phish-tank & 98.3\\
\hline
10 & TWSVM[5] & Alexa,Phish-tank &  98.05\\
\hline
11 &  Random Forest [4] &Common-Crawl,Alexa,Phish-tank & 94.26\\
\hline
12 & Multilayer Perceptron [13]& Kaggle & 98.4\\
\hline
13 & CNN+LSTM [33] & Phishtank,Common-Crawl & 93.28 \\
\hline
14 &  Random Forest with NLP based features[34]& Phish-tank & 97.99 \\
\hline
15 & Adaboost + SVM [21]& UCI Machine Learning Repository & 97.61 \\
\hline
16 & CNN-MHSA [24] &Phish-tank,5000 Best Websites & 99.84 \\
\hline
17 & Multilayer Perceptron[22]&UCI Machine Learning Repository,Kaggle & 96.65\\
\hline
18 & HNB+J48[28] & UCI Machine Learning Repository & 96.25\\
\hline
\end{tabular}
\end{table*}
It has the details of the sources of the datasets taken for experimenting, number of instances of website URLs taken to train model which can predict the phishing sites, along with the model which gave highest accuracy when compared with the other models. This data is taken from the papers collected.
\section{Challenges \& Discussion}
Some of the challenges are collected from the existing methods to improve their accuracy rate.\newline\newline
1. Reduce False positives\newline
In classification problem machine-learning gives a confusion matrix. In some classifiers, false positive rates are high i.e, even though the websites are legitimate the model classifies them as illegitimate sites, in this way end-users can't access the real website. If they can be reduced then the end-users can be able to access the legitimate sites without any problem.\newline\newline
2. Eliminate False negatives\newline
While predicting the accuracy, the classifiers give false negatives i.e, even though the websites are illegitimate the model classifies them as legitimate and this will result in damage, which includes loss of fame, corruption of systems and so on. These must be eliminated in order to prevent any harm to end-users and organisations.\newline\newline
3. Datasets and their modelling time\newline
Datasets play an important role in training models. By using old datasets the trained model may not be able to predict correctly as the model is unaware of new phishing attack vectors. Using current datasets is the solution. When the datasets are small, modelling time of the classifiers is not known as they will be trained in less time. When datasets differ in size their modelling time will alter so when we use larger datasets the modelling time will be known properly.\newline\newline
4. Selection and usage of features\newline
There are many features of a website such as URL, page, content features, domain features, source-code and so on which are used for detecting the sites. To decide what features can be used to train a model which can give more accuracy in detection is difficult. When only a single feature is used for detection then the prediction results may not be accurate. Using of multiple features of a website give more information about the site which can help in the detection process.\newline\newline
5. Sensitive words\newline
Use of sensitive words such as mail, bank, SMS and so on will have impact while predicting the sites. This may reflect in the results.\newline\newline
6. Embedded objects\newline
When a detection technique uses source code of a website for predicting the sites it extracts all the html tags but when there are any embedded objects such as i-frames, flash etc, it may be not able to detect properly.\newline\newline
7. Overfitting\newline
Overfitting happens when a model learns the detail and noise in the training data to the extent that it negatively impacts the performance of the model on new data. The problem is that these concepts do not apply to new data and negatively impact the models ability to generalize [78].
\section{Conclusion}We discussed phishing website detection techniques. The techniques include list-based, heuristic-based, visual-similarity and machine-learning. We compared the performance of different methods with respect to datasets. Future work includes, finding solutions for the challenges, so that any type of phishing sites can be detected as early as possible.

\section*{References}
[1]  A. A.A. and P. K., "Towards the Detection of Phishing Attacks," 2020 4th International Conference on Trends in Electronics and Informatics (ICOEI)(48184), Tirunelveli, India, 2020, pp. 337-343, doi: 10.1109/ICOEI48184.2020.9142967.\newline \newline
[2]C. Singh and Meenu, "Phishing Website Detection Based on Machine Learning: A Survey," 2020 6th International Conference on Advanced Computing and Communication Systems (ICACCS), Coimbatore, India, 2020, pp. 398-404, doi: 10.1109/ICACCS48705.2020.9074400. \newline \newline
[3]A. Al-Alyan and S. Al-Ahmadi, "Robust URL Phishing Detection Based on Deep Learning," KSII Transactions on Internet and Information Systems, vol. 14, no. 7, pp. 2752-2768, 2020. DOI: 10.3837/tiis.2020.07.001.\newline\newline
[4]	Rao, R.S., Vaishnavi, T. \& Pais, A.R. CatchPhish: detection of phishing websites by inspecting URLs. J Ambient Intell Human Comput 11, 813–825 (2020). https://doi.org/10.1007/s12652-019-01311-4 \newline \newline
[5]Rao, R.S., Pais, A.R. \& Anand, P. A heuristic technique to detect phishing websites using TWSVM classifier. Neural Comput \& Applic (2020). https://doi.org/10.1007/s00521-020-05354-z\newline \newline
[6]	J. Kumar, A. Santhanavijayan, B. Janet, B. Rajendran and B. S. Bindhumadhava, "Phishing Website Classification and Detection Using Machine Learning," 2020 International Conference on Computer Communication and Informatics (ICCCI), Coimbatore, India, 2020, pp. 1-6, doi: 10.1109/ICCCI48352.2020.9104161. \newline \newline
[7]M. Korkmaz, O. K. Sahingoz and B. Diri, "Feature Selections for the Classification of Webpages to Detect Phishing Attacks: A Survey," 2020 International Congress on Human-Computer Interaction, Optimization and Robotic Applications (HORA), Ankara, Turkey, 2020, pp. 1-9, doi: 10.1109/HORA49412.2020.9152934.\newline \newline
[8]	Aljofey, A.; Jiang, Q.; Qu, Q.; Huang, M.; Niyigena, J.-P. An Effective Phishing Detection Model Based on Character Level Convolutional Neural Network from URL. Electronics 2020, 9, 1514. https://doi.org/10.3390/electronics9091514 \newline \newline
[9]	M. N. Alam, D. Sarma, F. F. Lima, I. Saha, R. -E. -. Ulfath and S. Hossain, "Phishing Attacks Detection using Machine Learning Approach," 2020 Third International Conference on Smart Systems and Inventive Technology (ICSSIT), Tirunelveli, India, 2020, pp. 1173-1179, doi: 10.1109/ICSSIT48917.2020.9214225.\newline\newline
[10]S. Y. Yerima and M. K. Alzaylaee, "High Accuracy Phishing Detection Based on Convolutional Neural Networks," 2020 3rd International Conference on Computer Applications \& Information Security (ICCAIS), Riyadh, Saudi Arabia, 2020, pp. 1-6, doi: 10.1109/ICCAIS48893.2020.9096869. \newline \newline
[11]Adebowale, M.A., Lwin, K.T. and Hossain, M.A. (2020), "Intelligent phishing detection scheme using deep learning algorithms", Journal of Enterprise Information Management, Vol. ahead-of-print No. ahead-of-print. https://doi.org/10.1108/JEIM-01-2020-0036 \newline\newline
[12]Carlo Marcelo Revoredo da Silva, Eduardo Luzeiro Feitosa, Vinicius Cardoso Garcia, Heuristic-based strategy for Phishing prediction: A survey of URL-based approach,Computers |\& Security,Volume 88,2020,101613,ISSN 0167-4048,
https://doi.org/10.1016/j.cose.2019.101613.\newline\newline
[13]I. Saha, D. Sarma, R. J. Chakma, M. N. Alam, A. Sultana and S. Hossain, "Phishing Attacks Detection using Deep Learning Approach," 2020 Third International Conference on Smart Systems and Inventive Technology (ICSSIT), Tirunelveli, India, 2020, pp. 1180-1185, doi: 10.1109/ICSSIT48917.2020.9214132.\newline\newline
[14]F. Tajaddodianfar, J. W. Stokes and A. Gururajan, "Texception: A Character/Word-Level Deep Learning Model for Phishing URL Detection," ICASSP 2020 - 2020 IEEE International Conference on Acoustics, Speech and Signal Processing (ICASSP), Barcelona, Spain, 2020, pp. 2857-2861, doi: 10.1109/ICASSP40776.2020.9053670.\newline\newline
[15]M. Arivukarasi and A. Antonidoss, "Performance Analysis of Malicious URL Detection by using RNN and LSTM," 2020 Fourth International Conference on Computing Methodologies and Communication (ICCMC), Erode, India, 2020, pp. 454-458, doi: 10.1109/ICCMC48092.2020.ICCMC-00085.\newline\newline
[16]M. Korkmaz, O. K. Sahingoz and B. Diri, "Detection of Phishing Websites by Using Machine Learning-Based URL Analysis," 2020 11th International Conference on Computing, Communication and Networking Technologies (ICCCNT), Kharagpur, India, 2020, pp. 1-7, doi: 10.1109/ICCCNT49239.2020.9225561.\newline\newline
[17]N. Al-Milli and B. H. Hammo, "A Convolutional Neural Network Model to Detect Illegitimate URLs," 2020 11th International Conference on Information and Communication Systems (ICICS), Irbid, Jordan, 2020, pp. 220-225, doi: 10.1109/ICICS49469.2020.239536.\newline\newline
[18]Y. Su, "Research on Website Phishing Detection Based on LSTM RNN," 2020 IEEE 4th Information Technology, Networking, Electronic and Automation Control Conference (ITNEC), Chongqing, China, 2020, pp. 284-288, doi: 10.1109/ITNEC48623.2020.9084799.\newline\newline
[19]C. Opara, B. Wei and Y. Chen, "HTMLPhish: Enabling Phishing Web Page Detection by Applying Deep Learning Techniques on HTML Analysis," 2020 International Joint Conference on Neural Networks (IJCNN), Glasgow, United Kingdom, 2020, pp. 1-8, doi: 10.1109/IJCNN48605.2020.9207707.\newline\newline
[20]S. Priya, S. Selvakumar and R. L. Velusamy, "Detection of Phishing Attacks using Radial Basis Function Network Trained for Categorical Attributes," 2020 11th International Conference on Computing, Communication and Networking Technologies (ICCCNT), Kharagpur, India, 2020, pp. 1-6, doi: 10.1109/ICCCNT49239.2020.9225549.\newline\newline
[21]Abdulhamit Subasi, Emir Kremic,Comparison of Adaboost with MultiBoosting for Phishing Website Detection,Procedia Computer Science,Volume 168,2020,Pages 272-278,ISSN 1877-0509,https://doi.org/10.1016/j.procs.2020.02.251.\newline\newline
[22]Al-Ahmadi, Saad, PDMLP: Phishing Detection Using Multilayer Perceptron (2020). International Journal of Network Security \& Its Applications (IJNSA) Vol. 12, No.3, May 2020, Available at SSRN: https://ssrn.com/abstract=3624621\newline\newline
[23]Kumar, M.S., Indrani, B. Frequent rule reduction for phishing URL classification using fuzzy deep neural network model. Iran J Comput Sci (2020). https://doi.org/10.1007/s42044-020-00067-x \newline\newline
[24]Xi Xiao, Dianyan Zhang, Guangwu Hu, Yong Jiang, Shutao Xia,CNN–MHSA: A Convolutional Neural Network and multi-head self-attention combined approach for detecting phishing websites,Neural Networks,Volume 125,2020,Pages 303-312,ISSN 0893-6080,https://doi.org/10.1016/j.neunet.2020.02.013.\newline\newline
[25]Somesha, M., Pais, A.R., Rao, R.S. et al. Efficient deep learning techniques for the detection of phishing websites. Sādhanā 45, 165 (2020). https://doi.org/10.1007/s12046-020-01392-4 \newline\newline
[26]R. Purwanto, A. Pal, A. Blair and S. Jha, "PhishZip: A New Compression-based Algorithm for Detecting Phishing Websites," 2020 IEEE Conference on Communications and Network Security (CNS), Avignon, France, 2020, pp. 1-9, doi: 10.1109/CNS48642.2020.9162211.\newline\newline
[27]Sahar Abdelnabi, Katharina Krombholz, and Mario Fritz. 2020. VisualPhishNet: Zero-Day Phishing Website Detection by Visual Similarity. In Proceedings of the 2020 ACM SIGSAC Conference on Computer and Communications Security (CCS '20). Association for Computing Machinery, New York, NY, USA, 1681–1698. DOI:https://doi.org/10.1145/3372297.3417233\newline\newline
[28]Deep, Shekh Minhaz \& Zaman, Shihabuz \& Kawsar, Zul \& Ashaduzzaman, Md \& Pritom, Ahmed. (2019). Phishing Website Detection Using Effective Classifiers and Feature Selection Techniques. 10.13140/RG.2.2.24043.08483.\newline\newline
[29]Verma M.K., Yadav S., Goyal B.K., Prasad B.R., Agarawal S. (2019) Phishing Website Detection Using Neural Network and Deep Belief Network. In: Sa P., Bakshi S., Hatzilygeroudis I., Sahoo M. (eds) Recent Findings in Intelligent Computing Techniques. Advances in Intelligent Systems and Computing, vol 707. Springer, Singapore. https://doi.org/10.1007/978-981-10-8639-7\_30\newline\newline
[30]Feng, J. et al. “A Phishing Webpage Detection Method Based on Stacked Autoencoder and Correlation Coefficients.” J. Comput. Inf. Technol. 27 (2019): 41-54.\newline\newline
[31]Y. Huang, J. Qin and W. Wen, "Phishing URL Detection Via Capsule-Based Neural Network," 2019 IEEE 13th International Conference on Anti-counterfeiting, Security, and Identification (ASID), Xiamen, China, 2019, pp. 22-26, doi: 10.1109/ICASID.2019.8925000.\newline\newline
[32]N. N. Gana and S. M. Abdulhamid, "Machine Learning Classification Algorithms for Phishing Detection: A Comparative Appraisal and Analysis," 2019 2nd International Conference of the IEEE Nigeria Computer Chapter (NigeriaComputConf), Zaria, Nigeria, 2019, pp. 1-8, doi: 10.1109/NigeriaComputConf45974.2019.8949632.\newline\newline
[33]M. A. Adebowale, K. T. Lwin and M. A. Hossain, "Deep Learning with Convolutional Neural Network and Long Short-Term Memory for Phishing Detection," 2019 13th International Conference on Software, Knowledge, Information Management and Applications (SKIMA), Island of Ulkulhas, Maldives, 2019, pp. 1-8, doi: 10.1109/SKIMA47702.2019.8982427.\newline\newline
[34]V. M. Yazhmozhi and B. Janet, "Natural language processing and Machine learning based phishing website detection system," 2019 Third International conference on I-SMAC (IoT in Social, Mobile, Analytics and Cloud) (I-SMAC), Palladam, India, 2019, pp. 336-340, doi: 10.1109/I-SMAC47947.2019.9032492.\newline\newline
[35]K. Althobaiti, G. Rummani and K. Vaniea, "A Review of Human- and Computer-Facing URL Phishing Features," 2019 IEEE European Symposium on Security and Privacy Workshops (EuroS\&PW), Stockholm, Sweden, 2019, pp. 182-191, doi: 10.1109/EuroSPW.2019.00027. \newline \newline
[36]G. J. W. Kathrine, P. M. Praise, A. A. Rose and E. C. Kalaivani, "Variants of phishing attacks and their detection techniques," 2019 3rd International Conference on Trends in Electronics and Informatics (ICOEI), Tirunelveli, India, 2019, pp. 255-259, doi: 10.1109/ICOEI.2019.8862697. \newline \newline
[37]A. Nakamura and F. Dobashit, "Proactive Phishing Sites Detection," 2019 IEEE/WIC/ACM International Conference on Web Intelligence (WI), Thessaloniki, Greece, 2019, pp. 443-448.\newline \newline
[38]M. Zabihimayvan and D. Doran, "Fuzzy Rough Set Feature Selection to Enhance Phishing Attack Detection," 2019 IEEE International Conference on Fuzzy Systems (FUZZ-IEEE), New Orleans, LA, USA, 2019, pp. 1-6, doi: 10.1109/FUZZ-IEEE.2019.8858884. \newline \newline
[39]M. Chatterjee and A. Namin, "Detecting Phishing Websites through Deep Reinforcement Learning," 2019 IEEE 43rd Annual Computer Software and Applications Conference (COMPSAC), Milwaukee, WI, USA, 2019, pp. 227-232, doi: 10.1109/COMPSAC.2019.10211. \newline \newline
[40]E. Zhu, Y. Chen, C. Ye, X. Li and F. Liu, "OFS-NN: An Effective Phishing Websites Detection Model Based on Optimal Feature Selection and Neural Network," in IEEE Access, vol. 7, pp. 73271-73284, 2019, doi: 10.1109/ACCESS.2019.2920655. \newline \newline
[41]Y. Huang, Q. Yang, J. Qin and W. Wen, "Phishing URL Detection via CNN and Attention-Based Hierarchical RNN," 2019 18th IEEE International Conference On Trust, Security And Privacy In Computing And Communications/13th IEEE International Conference On Big Data Science And Engineering (TrustCom/BigDataSE), Rotorua, New Zealand, 2019, pp. 112-119, doi: 10.1109/TrustCom/BigDataSE.2019.00024.\newline \newline
[42]Rao, R.S., Pais, A.R. Detection of phishing websites using an efficient feature-based machine learning framework. Neural Comput \& Applic 31, 3851–3873 (2019). https://doi.org/10.1007/s00521-017-3305-0 \newline \newline
[43]S. Roopak, A. P. Vijayaraghavan and T. Thomas, "On Effectiveness of Source Code and SSL Based Features for Phishing Website Detection," 2019 1st International Conference on Advanced Technologies in Intelligent Control, Environment, Computing \& Communication Engineering (ICATIECE), Bangalore, India, 2019, pp. 172-175, doi: 10.1109/ICATIECE45860.2019.9063824.\newline \newline
[44]H. Chapla, R. Kotak and M. Joiser, "A Machine Learning Approach for URL Based Web Phishing Using Fuzzy Logic as Classifier," 2019 International Conference on Communication and Electronics Systems (ICCES), Coimbatore, India, 2019, pp. 383-388, doi: 10.1109/ICCES45898.2019.9002145.\newline \newline
[45]M. M. Yadollahi, F. Shoeleh, E. Serkani, A. Madani and H. Gharaee, "An Adaptive Machine Learning Based Approach for Phishing Detection Using Hybrid Features," 2019 5th International Conference on Web Research (ICWR), Tehran, Iran, 2019, pp. 281-286, doi: 10.1109/ICWR.2019.8765265.\newline \newline 
[46]A. F. Nugraha and L. Rahman, "Meta-Algorithms for Improving Classification Performance in the Web-phishing Detection Process," 2019 4th International Conference on Information Technology, Information Systems and Electrical Engineering (ICITISEE), Yogyakarta, Indonesia, 2019, pp. 271-275, doi: 10.1109/ICITISEE48480.2019.9003952.\newline \newline
[47]T. M. Abed and H. B. Abdul-Wahab, "Anti-Phishing System Using Intelligent Techniques," 2019 2nd Scientific Conference of Computer Sciences (SCCS), Baghdad, Iraq, 2019, pp. 44-50, doi: 10.1109/SCCS.2019.8852601.\newline \newline
[48]Jain, A.K., Gupta, B.B. A machine learning based approach for phishing detection using hyperlinks information. J Ambient Intell Human Comput 10, 2015–2028 (2019). https://doi.org/10.1007/s12652-018-0798-z\newline \newline
[49]Sahingoz, O.K., Baykal, S.I., Bulut, D., Phishing detection from
urls by using neural networks\newline\newline
[50]S. Le Page, G. Jourdan, G. V. Bochmann, J. Flood and I. Onut, "Using URL shorteners to compare phishing and malware attacks," 2018 APWG Symposium on Electronic Crime Research (eCrime), San Diego, CA, 2018, pp. 1-13, doi: 10.1109/ECRIME.2018.8376215\newline\newline
[51]A. Niakanlahiji, B. Chu and E. Al-Shaer, "PhishMon: A Machine Learning Framework for Detecting Phishing Webpages," 2018 IEEE International Conference on Intelligence and Security Informatics (ISI), Miami, FL, 2018, pp. 220-225, doi: 10.1109/ISI.2018.8587410. \newline \newline
[52] Doyen Sahoo, Chenghao Liu, and Steven C H Hoi. 2017. Malicious URL detection using machine learning: A survey. arXiv preprint arXiv:1701.07179 (2017)\newline\newline
[53]H. Shirazi, K. Haefner and I. Ray, "Fresh-Phish: A Framework for Auto-Detection of Phishing Websites," 2017 IEEE International Conference on Information Reuse and Integration (IRI), San Diego, CA, 2017, pp. 137-143, doi: 10.1109/IRI.2017.40. \newline \newline
[54]S. Haruta, H. Asahina and I. Sasase, "Visual Similarity-Based Phishing Detection Scheme Using Image and CSS with Target Website Finder," GLOBECOM 2017 - 2017 IEEE Global Communications Conference, Singapore, 2017, pp. 1-6, doi: 10.1109/GLOCOM.2017.8254506.\newline \newline
[55]Heartfield R, Loukas G. A taxonomy of attacks and a survey of defence mechanisms for semantic social engineering attacks. ACM Computing Surveys (CSUR). 2015 Dec 9;48(3):1-39.\newline \newline
[56]https://www.itgovernance.eu/blog/en/the-5-most-common-types-of-phishing-attack\newline \newline
[57]https://searchsecurity.techtarget.com/definition
/whaling\newline \newline
[58]https://cyware.com/news/smishing-and-vishing-whats-the-difference-between-them-4f55d408/\newline \newline
[59]https://www.imperva.com/learn/application-security/phishing-attack-scam/\newline \newline
[60]https://www.pcworld.com/article/135293/article.html\newline\newline
[61]http://www2.deloitte.com/content/dam/Deloitte/sg/Documents/risk/sea-risk-cyber-101-part10.pdf\newline\newline
[62]https://www.alexa.com/topsites\newline\newline
[63]http://index.commoncrawl.org/\newline\newline
[64]https://www.phishtank.com/developer\_info.php\newline\newline
[65]https://openphish.com/\newline\newline
[66]https://archive.ics.uci.edu/ml/datasets/phishing+websites\newline\newline
[67]https://majestic.com/reports/majestic-million\newline\newline
[68]https://github.com/ebubekirbbr/pdd/tree/master/input \newline\newline
[69]https://www.ibm.com/cloud/blog/ai-vs-machine-learning-vs-deep-learning-vs-neural-networks\newline\newline
[70]https://linuxhint.com/kali-linux-set/\newline\newline
[71]https://en.wikipedia.org/wiki/Feedforward\_neural\_network \newline\newline
[72]https://en.wikipedia.org/wiki/Recurrent\_neural\_network\newline\newline
[73]https://en.wikipedia.org/wiki/Capsule\_neural\_network\newline\newline
[74]https://en.wikipedia.org/wiki/Long\_short-term\_memory\newline\newline
[75]https://www.kaggle.com/datasets\newline\newline
[76]http://5000best.com/websites/\newline\newline
[77]https://www.kdnuggets.com/2020/02/deep-neural-networks.html\newline\newline
[78]https://machinelearningmastery.com/overfitting-and-underfitting-with-machine-learning-algorithms/\newline\newline
\end{document}